\documentclass{iaus}
\usepackage{graphicx}

\title[Eclipsing binaries in the MACHO database] 
{Eclipsing binaries in the LMC: a wealth of data for astrophysical tests}

\author[A. Derekas {\it et al.}]   
{A. Derekas, L. L. Kiss \and T. R. Bedding}

\affiliation{School of Physics, University of Sydney, NSW 2006, Australia
\break email: derekas@physics.usyd.edu.au}

\pubyear{2007}
\volume{240}  
\pagerange{--}
\date{?? and in revised form ??}
\jname{Binary Stars as Critical Tools \& Tests \\ in Contemporary Astrophysics}
\editors{W. I. Hartkopf, E. F. Guinan \& P. Harmanec, eds}
\begin{document}

\maketitle

\begin{abstract}

We have analysed publicly available MACHO observations of 6833 variable stars 
in the Large Magellanic Cloud, classified as eclipsing binaries. After finding 
that a significant fraction of the sample was misclassified, we redetermined
periods and variability class for all stars, producing a clean sample of 3031
eclipsing binaries. We have investigated their distribution in the 
period-color-luminosity space, which was used, for example, to assign a 
foreground probability to every object and establish new period-luminosity 
relations to selected types of eclipsing stars. We found that the orbital 
period distribution of LMC binaries is very similar to those of the SMC and the
Milky Way. We have also determined the rate of period change for every star
using the O--C method, discovering about 40 eclipsing binaries 
with apsidal motion, 45 systems with cyclic period changes and about 80 stars
with parabolic O--C diagrams. In a few objects we discovered gradual 
amplitude variation, which can be explained by changes in 
the orbital inclination caused by a perturbing third body in the system.


\end{abstract}

\firstsection 

\section{Introduction}

The last decade witnessed the birth of a new research field, the large-scale
study of variable stars in external galaxies. This has first been made 
possible by the huge databases of microlensing observations of the Magellanic
Clouds, like the MACHO, OGLE and EROS projects (see Szabados \& Kurtz 2000 for
reviews). These programs (beyond their primary purpose) resulted in the
discovery of thousands of new eclipsing binaries with an unprecedented
homogeneous coverage of their light curves opening a new avenue of the binary
star research. Here we present the first results of an analysis of the publicly
available MACHO light curves. The main aim of the project is to measure period
changes and discover eclipsing binaries with pulsating components.

\section{General properties of the sample}

We have analysed MACHO lights curves for 6833 stars that were originally
classified as eclipsing binaries. After re-determining the period and
classifying the stars based on their light curve shapes, only 3031 stars 
remained as genuine eclipsing or ellipsoidal variables (the rest being
Cepheids, RR Lyraes and other non-eclipsing variable stars).  The period
distribution of this binary sample is bimodal, with the strongest 
peak between 1 and 2 days. Roughly 20\% of stars have periods
longer than 10 days; many of them show W UMa-like light curve shapes,
suggesting ellipsoidal variability of giant componens.

We classified the binary sample using Fourier-decomposition of the phase
diagrams. Two coefficients, ${\rm a_{2}}$ and ${\rm a_{4}}$, of the cosine
decomposition ${\rm \sum_{i=1}^4 a_{i} \cos (2\pi i \varphi)}$ allow a
well-defined distinction between detached, semi-detached and contact binaries
(Pojma\'nski 2002). The results show that the sample is dominated by bright
main-sequence detached (50\%) and semi-detached (30\%) binaries. Contact
systems comprise 20\% of the sample; the short period systems are all
foreground objects in the Milky Way, while longer periods belong to red giant
binaries.

We used the Color-Magnitude Diagram (CMD) to clean the sample of the foreground
objects. For this we took evolutionary models of Castellani et al. (2003) and
calculated the locations of certain minimal orbital period values (where two
identical model stars are in contact). The cleaned sample contains about 2800
LMC binaries.

Detached and semi-detached binaries are spread uniformly in the period-K
magnitude plane, while there is a well-defined sequence for the contact
systems. We found that the widely accepted sequence of eclisping binaries
between Seqs. C and D, known as Seq. E, does not exist. The correct position
for Seq. E is at periods a factor of two greater. A simple Roche-model
describes Seq. E very well. Although Seq. E seems to merge into Seq. D of the
Long Secondary Periods (Wood et al. 2004), the two groups are significantly
different in their color and amplitude properties (Derekas et al. 2006).

\begin{figure}
\begin{center}
\includegraphics[width=8cm]{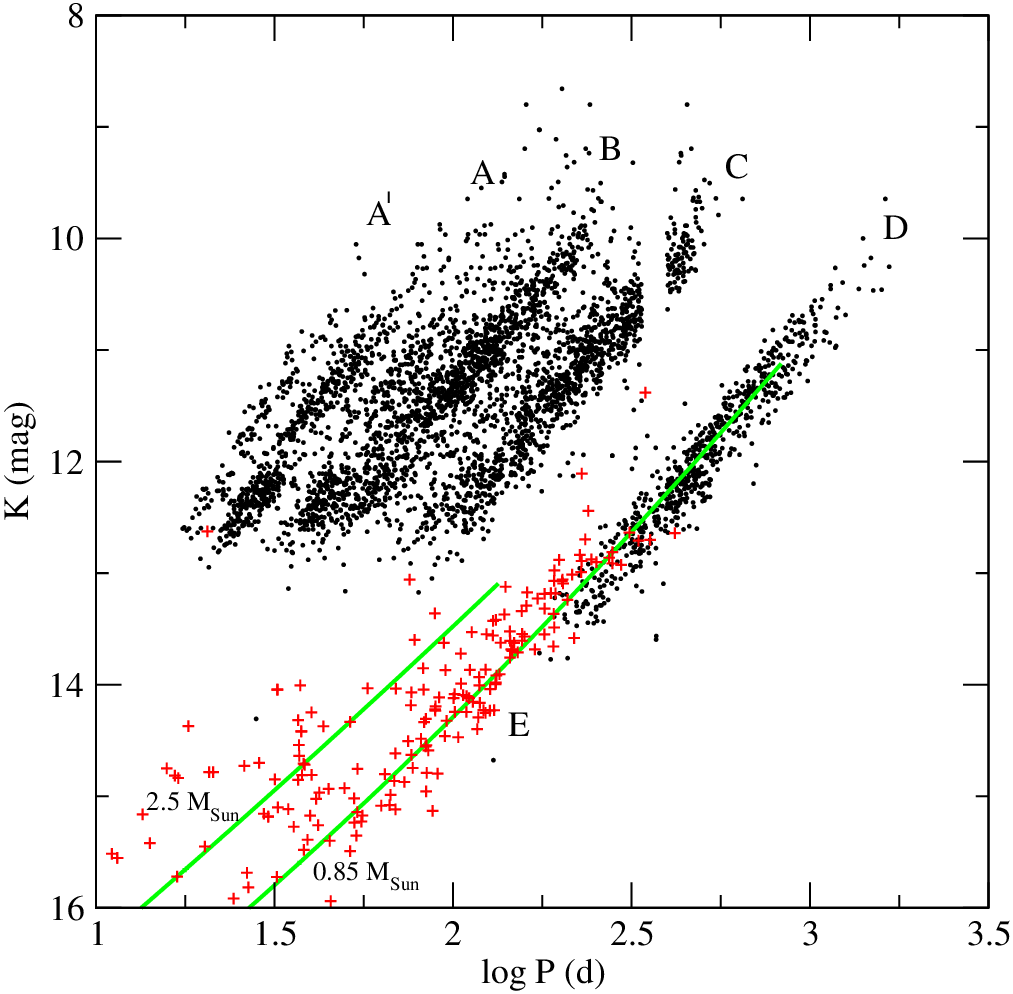}
  \caption{P--L relations of ellipsoidal variables (red  pluses) red
       giant pulsators (black dots). The two lines show a simple model using
       evolutionary tracks and Roche-geometry.}
\end{center}
\end{figure}

\section{Period changes and secular amplitude variations}

From the 8 years of MACHO observations we measured period changes using the
O--C method applied to seasonal subsets of the data. We found about 80
parabolic and 45 cyclic period changes, the rest showing linear O--C diagrams.
A significant fraction of the former two groups are candidates for light-time
effect in hierarchic triple systems. One example is shown in the top panels of
Fig. 2. In about 40 eccentric binaries we measured different O--C variations for the
primary and the secondary minima, which indicates apsidal motion (bottom panels
in Fig.  2). With this we double the number of known binaries with apsidal
motion in the LMC (Michalska \& Pigulski 2005).

\begin{figure}
\includegraphics[width=13cm]{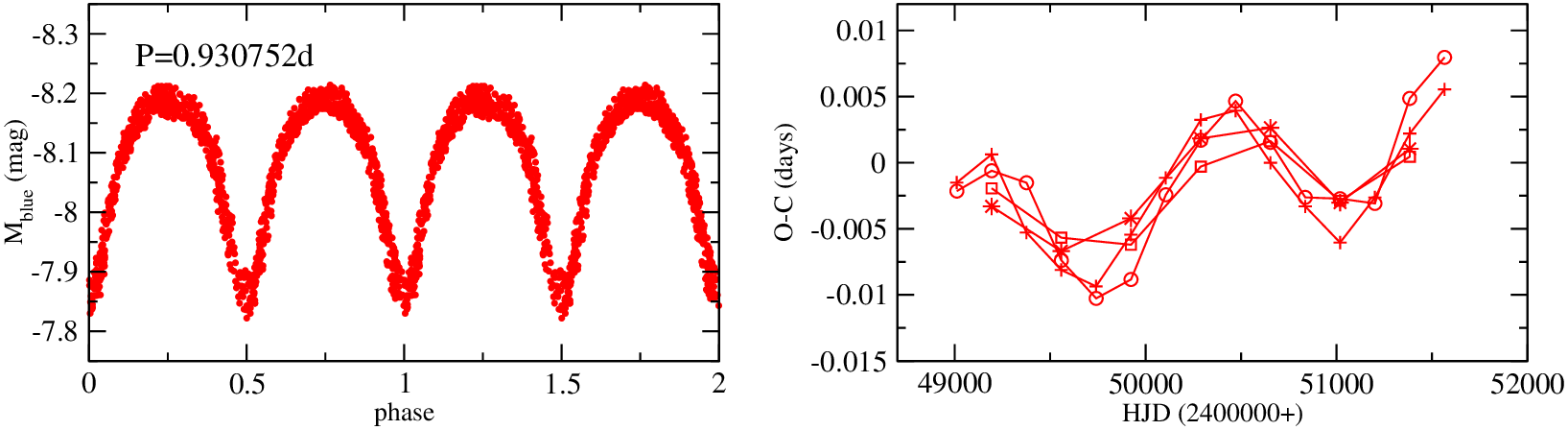}
\vskip2mm
\includegraphics[width=13cm]{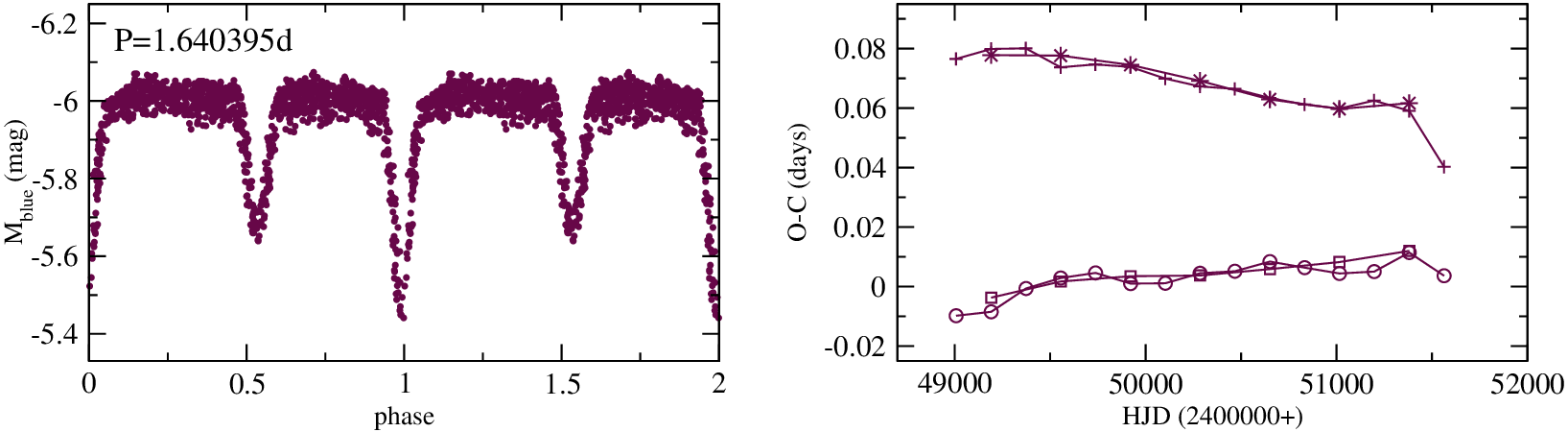}
  \caption{Examples for cyclic O--C diagram and apsidal motion.}
\end{figure}

In a few objects we discovered gradual amplitude variation, which can be
explained by rapid variations in the orbital geometry, most likely in
inclination. A third body in the system can perturb the eclipsing pair in such
a way that the eclipse depth, as a sensitive indicator of the inclination
variations, follows these perturbations. In Fig. 3 we present an example ($\rm
P_{ecl}=0.77d$), for which the large scatter in the O--C diagram may suggest an
orbital period of $\leq 100d$ for the hypothetical third companion.

\begin{figure}
\includegraphics[width=13cm]{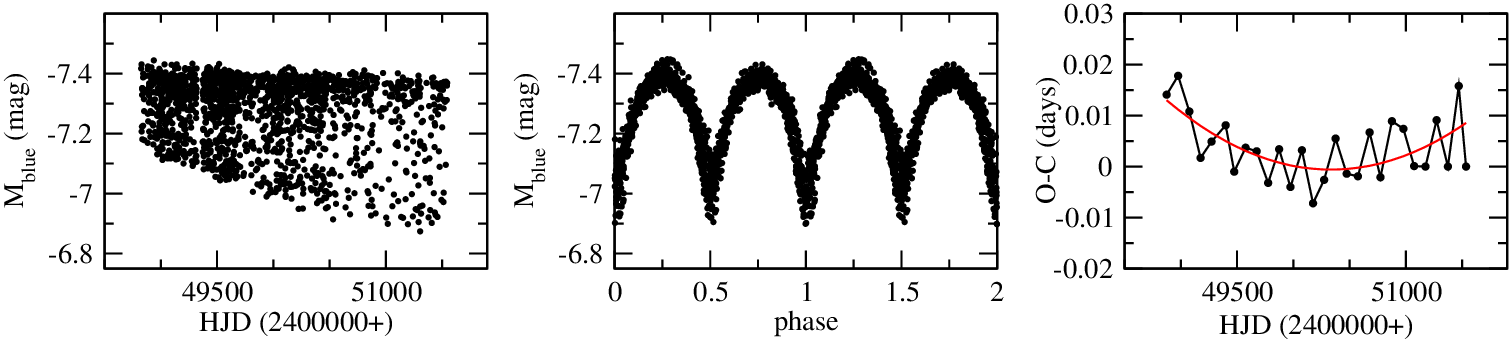}
  \caption{An example for changing minimum depth.}
\end{figure}

\begin{acknowledgments}
\begin{footnotesize} 

This work has been supported by the Australian Research Council. LLK is
supported by a University of Sydney Postdoctoral Research Fellowship. AD is
supported by an Australian Postgraduate Research Award.  This paper utilizes
public domain data obtained by the MACHO Project, jointly funded by the US
Department of Energy through the University of California, Lawrence Livermore
National Laboratory under contract No. W-7405-Eng-48, by the National Science
Foundation through the Center for Particle Astrophysics of the University of
California under cooperative agreement AST-8809616, and by the Mount Stromlo
and Siding Spring Observatory, part of the Australian National University.

\end{footnotesize}

\end{acknowledgments}


\begin{thebibliography}{}

\item Castellani, V., et al., 2003, A\&A, 404, 645

\item Derekas et al., 2006, ApJ, 650, L55

\item Michalska, G., Pigulski, A., 2005, A\&A, 434, 89

\item Pojma\'nski, G., 2002, Acta Astron., 52, 397

\item Szabados, L., Kurtz, D.W., (Eds.), 2000, ASP Conf. Series, Vol. 203

\item Wood P. R., Olivier, E. A., \& Kawaler, S. D., 2004, ApJ, 604, 800




\end{thebibliography}
\end{document}